# *In vivo* deep tissue imaging using wavefront shaping optical coherence tomography

Hyeonseung Yu,[a, *] Jaehyun Peter Lee,[b, *] KyeoReh Lee,[a] Jaeduck Jang,[c] Jaeguyn Lim,[d] Wooyoung Jang,[d] Yong Jeong,[b,+] YongKeun Park [a,+]

[a] Department of Physics, Korea Advanced Institute of Science and Technology, Daejeon 34141, South Korea
[b] Department of Bio and Brain Engineering, Korea Advanced Institute of Science and Technology, Daejeon 34141, South Korea
[c] Samsung Advanced Institute of Technology, Yongin, Gyeonggi 16677, South Korea
[d] Samsung Electronics, Suwon, Gyeonggi 16677, South Korea
*These authors contributed equally to this work
+Address all correspondence to:
YongKeun Park, E-mail: yk.park@kaist.ac.kr or Yong Jeong, E-mail: yong@kaist.ac.kr

**Abstract**. Multiple light scattering in tissue limits the penetration of optical coherence tomography (OCT) imaging. Here, we present *in-vivo* OCT imaging of a live mouse using wavefront shaping to enhance the penetration depth. A digital micro-mirror device (DMD) was used in a spectral-domain OCT system for complex wavefront shaping of an incident beam which resulted in the optimal delivery of light energy into deep tissue. *Ex-vivo* imaging of chicken breasts and mouse ear tissues showed enhancements in the strength of the image signals and the penetration depth, and *in-vivo* imaging of the tail of a live mouse provided a multilayered structure inside the tissue, otherwise invisible in conventional OCT imaging. Signal enhancements by a factor of 3–7 were acquired for various experimental conditions and samples.

Optical coherence tomography (OCT) provides non-invasive imaging of biological tissues based on low coherence interferometry[1]. Analogous to ultrasound imaging, the use of infrared wavelengths in OCT provides a micrometer resolution and has successfully been used in various applications, particularly in ophthalmology. However, when imaging general tissues such as skin and internal organs, the penetration depth of OCT does not exceed a few millimeters due to multiple scattering of light caused by inhomogeneous distributions of refractive indexes in the tissue. Because the imaging contrast in OCT is from single back scattering from a scatter, the optical signal decays exponentially as a function of depth in the presence of multiple scatterings[2]. Beyond the one transport mean free path which corresponds to 1–2 mm in most biological tissues[3], a single back-scattered signal is severely degraded, and multiple scattering becomes dominant which limits the penetration depth of OCT.

Due to this limited penetration depth in OCT, imaging highly scattering tissues, such as breast and skin tissues, has not been fully explored, whereas retina imaging using OCT in ophthalmology has been extensively investigated[4,5]. This limited penetration depth of OCT is unfortunate because OCT has much to offer to various fields in biology/biotechnology/life sciences and medical diagnoses with its unique non-invasive optical sectioning capability, its high spatial resolving power, and its lack of radiation damage. For example, pathological studies in dermatology have shown the potential of OCT[6] while the accessible depth is restricted to the epidermis and upper dermis layers[7]. OCT can also be useful for non-invasive diagnoses of cancers[8]; however, the imaging depth of OCT is confined to a superficial layer which prevents early cancer detection. Multimodal OCT systems equipped with photoacoustic microscopy have been proposed to provide microvascular imaging at greater depths[9-11] while morphological information on deep tissue regions, required for the exact understanding of vascular properties, is still inaccessible.

Several approaches have been proposed to suppress multiple light scattering in OCT imaging. Refractive index matching using optical clearing agents can reduced multiple scatterings[12], but it requires a long setting time (~10 min.). Adaptive optics approaches have been used to correct the aberration[13-16]; however, they mainly address aberrations but do not suppress multiple scatterings due to the limited degree of control. Spatial and frequency compounding methods reduce speckle noises caused by multiple scatterings and provide an enhanced signal-to-noise ratio (SNR)[17-20]. However, the penetration depth is yet to be improved in spatial and frequency compounding methods.

Recently, wavefront shaping techniques have shown potential for enhancing the penetration depth in low-coherence reflection imaging. By controlling the wavefront of an impinging beam, control of the light field transmitted through a turbid media can be achieved through linear coherent relationships between input and transmitted fields, which are described by scattering matrices[21-23]. Wavefront shaping approaches have been applied to low-coherence interferometry, including selective focusing in optical coherence microscopy[24]; the penetration depth has been enhanced in spectral domain OCT[25, 26], called wavefront shaping OCT (WS-OCT); imaging deep in turbid media by collective accumulation of single-scattered waves from a time-resolved reflection matrix[27]. Exploiting this approach, significant advances have been made in the last few years in enhancing the penetration depth of low coherence interferometric imaging systems. However, only demonstrations with static phantoms have been reported. Imaging biological tissues *in vivo* with wavefront shaping approaches still remains unexplored in OCT systems mainly because (1) it takes a long time to acquire the information about the scattering matrices and control the wavefront of the impinging beams and (2) the movements in tissues scramble the information scattering matrixes.

In this study, we report the *in-vivo* imaging of a live mouse using WS-OCT and demonstrate enhancements in both the signal-to-noise ratio (SNR) and penetration depth. A digital micro-mirror device (DMD) was used in a spectral domain OCT system to control the wavefront of the input beams. Significant enhancements in the SNR and penetration depth were observed for an *ex vivo* chicken breast, *ex vivo* mouse ear, and *in vivo* mouse tail. Compared to an uncontrolled beam illumination, the image signals were enhanced by a factor of 3–7 depending on the types of samples and depth positions. In particular, multiple layer structures in the mouse tail were clearly resolved with WS-OCT; otherwise, only the top layer is visible with a conventional OCT approach.

The concept of WS-OCT is shown in Fig. 1. In conventional OCT, a Gaussian beam impinges onto a sample [Fig. 1(a)]. However, multiple light scattering in the tissue prevents optimal light delivery deep into the tissue; highly inhomogeneous distributions of refractive indices result in the scrambling of the optical paths and the formation of a speckle field inside the tissue. As a result, it prevents the delivery of light to a target depth as well as coherent gating in low-coherence interferometry. When the wavefront of an input beam is controlled and optimized, an optical focus at the target depth can be formed in the presence of multiple light scattering [Fig. 1(b)]. This can also be understood in a time reversal manner; when light is emitted deep inside a turbid medium, it will have a complex wavefront after transmitting through the medium, and when light is impinged in a reversed direction with this complex wavefront, a focus can be generated in the original location. To find the optimized wavefront which enhanced the delivery of light to the target scatter, for each A-line scan, we measured the reflection responses corresponding to various optical wavefront patterns of impinging beams. Then, from the measured reflection responses, the optimal wavefront can be obtained by finding the constructive interference condition[25, 26]. Finally, an illumination with the optimized wavefront is applied to the sample, and then, a depth-enhanced OCT image is obtained.

The experimental setup is shown in Fig. 1(c). The wavefront shaping method is applied to a spectral-domain OCT system. A DMD (0.7 XGA, 23 kHz frame rate, Texas Instruments, United States) is used for the modulation of an incident beam. A reference and a sample beam are coupled through a single mode fiber and then reaches a spectrometer consisting of a grating and a line CCD (SU1024-LDH2, 92 kHz frame rate, Sensors Unlimited Inc., United States). For a single A-scan optimization, input DMD pattern scans, 7,500 2-D Hadamard basis, and the corresponding A-scan OCT signals are obtained. Then, the optimal DMD patterns are calculated for each depth by finding the phase matching conditions from the measured responses. Next, the calculated optimal DMD patterns are projected to the sample sequentially, and the enhanced signals at each depth are compounded to produce a depth-enhanced A-scan image. By scanning over 15 A-scans, a 2-D enhanced image is acquired. The segment size of a pattern was set to 23×23 micro pixels in the DMD. The acquisition time for a single A-scan profile is 30 second. The principle of this method is analogues to finding a transmission matrix in turbid media[28-30]; however, the present method measured coherence- and pinhole-gated reflection responses.

To demonstrate the applicability to biological tissue, *ex vivo* chicken breast tissue was imaged with WS-OCT. The superficial layer of sliced chicken tissue was prepared and measured [Fig. 2(a)]. 2-D images of the tissue obtained with WS-OCT are shown in Fig. 2(b). The reduced scattering coefficient of the chicken tissue was measured to be 0.92 mm$^{-1}$ using an integrating sphere (UPK-100-F, Gigahertz-Optik) based on the inverse adding-doubling method [31]. For comparison purposes, the sample was also imaged with the controlled wavefront (Gaussian beam) and the spatial compounding method in Figs. 2(c) and (d), respectively. The image with the spatial compounding method was obtained by averaging 25 OCT images recorded with 25 random illumination patterns. The powers of the input beams were set to be the same (0.55 mW) for all three cases. It can clearly be seen that WS-OCT greatly enhances both the SNR and the penetration depth for *ex vivo* tissue compared with existing methods. In particular, WS-OCT enables the imaging of structures deep inside the tissue otherwise inaccessible with conventional approaches [the arrows in Fig. 2(b)].

To validate the tissue structures in the images obtained with WS-OCT, we conducted an experiment for the uncontrolled wavefront with double the incident power at 1.1 mW [Fig. 2(e)]. The structures obtained with WS-OCT with 0.55 mW is comparable to that obtained with the uncontrolled wavefront with 1.1 mW, and this shows that the enhanced signals in WS-OCT resulted from real scatters rather than false artifacts. For a quantitative comparison, the averaged depth profiles over 15 different A-scans in Figs. 2(b-e) are plotted in Fig. 2(f). The enhancement factor η in WS-OCT is defined as the ratio of the signal for the optimized wavefront to that for the uncontrolled wavefront with the same incident power. The enhancement factor at the first layer of the tissue along the green dashed-line was 6.36, showing a significant improvement over conventional OCT imaging.

Then, we imaged the *ex vivo* ear tissue of a mouse. A specific pathogen-free C57BL/6J mouse (Jackson Laboratory, U.S.A.) was anesthetized with intraperitoneal injection of a Tiletamine-Zolazepam and Xylazine mixture (30:10 mg/kg body weight), and a piece of an ear was dissected for the imaging shown in Fig. 2(g). All experimental procedures were approved by the Institutional Animal Care and Use Committee (IACUC) of Korea Advanced Institute of Science and Technology (KAIST). The reduced scattering coefficient of the ear tissue was 0.53 mm$^{-1}$. The experimental results for the wavefront shaping, the uncontrolled beam, the spatial compounding, and double-powered uncontrolled beam are, respectively, presented in Figs. 2(h-k). Similar to the case with the chicken breast tissue, the wavefront shaping shows an apparently stronger signal and deeper penetration depth compared to existing methods. In Fig. 2(l), the averaged depth profiles over 15 different A-scans are plotted for comparison purposes. An enhancement factor of 5.37 was obtained at the upper layer (indicated as the gray dashed line). From the chicken breast tissue and the mouse ear results, we can conclude that the wavefront shaping technique successfully works for *ex-vivo* biological tissues with enhanced SNR and penetration depth.

Next, *in vivo* images of a mouse tail were obtained with WS-OCT. A mouse of the same type used in the *ex-vivo* study was anesthetized using the same procedures. To minimize the movement of the sample, the tale was stably taped down on the sample stage. A photograph of the mouse under imaging and micrographs of a tail section slice of 30 μm after Haematoxylin and eosin (H&E) stain are shown in Fig. 3(a). After 20 min. from the anesthetic injection, three different locations on the distal end of the mouse tail were imaged. The reduced scattering coefficient of the tail tissue was measured as 1.212 mm$^{-1}$. Experimental results for the wavefront shaping, the uncontrolled beam, the spatial compounding method, and double-powered uncontrolled beam are shown in Fig. 3. For all locations, WS-OCT showed a significantly improved penetration depth and SNR. In particular, multiple layered structures were visualized with WS-OCT: the double layered structure at location 1 [Fig. 3(b)] and the triple layered structures at location 2 and 3 [Fig. 3(g) and 3(l)] as indicated by gray arrows (note the difference in scales between location 1 and 2, 3). The shapes and locations of these structures corresponds to epidermis, connective tissue textures, and sebaceous glands. However, the uncontrolled beam and the spatial compounding method only visualized the most superficial single layer at all locations. The average depth profiles for 15 A-scans are plotted for each location in Figs. 3(f, k, q). The enhancement factors for the first layer were 5.26, 6.88 and 6.06 at location 1, 2 and 3, respectively. These enhancement factors are comparable to that of the former *ex-vivo* experiments, implying that the efficiency of wavefront shaping is not severely degraded in *in vivo* environments. The enhancement factors at the second peaks were 4.47 for location 1 and 3.97 for location 2, slight smaller than those of the first layers. Notably, the enhancement factor decreases as the depth increases. This may be due to the faction of uncollected light will increase as the depth of the scatters increases. At location 3, although the intensity level of the second peak [Fig. 3(q), the orange arrow] is much smaller than that of the other locations, an enhancement factor of 3.36 is still comparable to the other locations which implies the enhancement does not depend on the absolute level of intensity.

In summary, we present *ex vivo* and *in vivo* tissue images with WS-OCT and demonstrated a significant enhancement in the penetration depth and SNR. For *in vivo* mouse tail imaging, the present approach unraveled the multilayered structures otherwise invisible with conventional OCT with Gaussian input beams. The signal enhancement factor compared to the uncontrolled input beam reached up to 6.88 at the superficial layer.

This result clearly suggests that wavefront shaping is a promising method of controlling multiple light scattering for *in vivo* imaging of highly turbid tissue. Yet, the current work only demonstrated the imaging of movement-free samples such as mouse tails, due to the slow acquisition speed (30 seconds for each A-scan measurement). The acquisition speed should be further enhanced so that the present approach can be applicable to the imaging of moving parts such as ears, feet, or living organs. Recent studies suggest that the acquisition time must be reduced to the order of a millisecond to overcome the de-correlation time of biological tissues[32-34]. We anticipate that the further improvement of the measurement speed of the present can be accomplished with a field programmable gate array (FPGA) processing a graphics process unit (GPU), and this set up will lead to widespread applications of wavefront shaping approaches in OCT. Furthermore, the present method can also be expanded to other OCT modalities such as polarization-sensitive or spectroscopic signals, by exploiting large degree of freedoms in multiple scattering[21, 35, 36].

References
1. D. Huang, E. A. Swanson, C. P. Lin, J. S. Schuman, W. G. Stinson, W. Chang, M. R. Hee, T. Flotte, K. Gregory, C. A. Puliafito and J. G. Fujimoto, "Optical Coherence Tomography," *Science* **254**(5035), 1178-1181 (1991)
2. A. F. Fercher, W. Drexler, C. K. Hitzenberger and T. Lasser, "Optical coherence tomography - principles and applications," *Rep Prog Phys* **66**(2), 239-303 (2003)
3. D. A. Boas, C. Pitris and N. Ramanujam, *Handbook of biomedical optics*, CRC Press, Boca Raton (2011).
4. M. L. Gabriele, G. Wollstein, H. Ishikawa, L. Kagemann, J. Xu, L. S. Folio and J. S. Schuman, "Optical coherence tomography: history, current status, and laboratory work," *Investigative ophthalmology & visual science* **52**(5), 2425 (2011)
5. B. E. Bouma, S. H. Yun, B. J. Vakoc, M. J. Suter and G. J. Tearney, "Fourier-domain optical coherence tomography: recent advances toward clinical utility," *Curr Opin Biotech* **20**(1), 111-118 (2009)
6. J. Welzel, E. Lankenau, R. Birngruber and R. Engelhardt, "Optical coherence tomography of the human skin," *Journal of the American Academy of Dermatology* **37**(6), 958-963 (1997)
7. T. Gambichler, V. Jaedicke and S. Terras, "Optical coherence tomography in dermatology: technical and clinical aspects," *Archives of Dermatological Research* **303**(7), 457-473 (2011)
8. B. J. Vakoc, D. Fukumura, R. K. Jain and B. E. Bouma, "Cancer imaging by optical coherence tomography: preclinical progress and clinical potential," *Nat Rev Cancer* **12**(5), 363-368 (2012)
9. S. Jiao, Z. Xie, H. F. Zhang and C. A. Puliafito, "Simultaneous multimodal imaging with integrated photoacoustic microscopy and optical coherence tomography," *Optics Letters* **34**(19), 2961-2963 (2009)
10. L. Li, K. Maslov, G. Ku and L. V. Wang, "Three-dimensional combined photoacoustic and optical coherence microscopy for in vivo microcirculation studies," *Optics Express* **17**(19), 16450-16455 (2009)
11. E. Z. Zhang, B. Povazay, J. Laufer, A. Alex, B. Hofer, B. Pedley, C. Glittenberg, B. Treeby, B. Cox and P. Beard, "Multimodal photoacoustic and optical coherence tomography scanner using an all optical detection scheme for 3D morphological skin imaging," *Biomed. Opt. Express* **2**(8), 2202-2215 (2011)
12. V. V. Tuchin and V. Tuchin, *Tissue optics: light scattering methods and instruments for medical diagnosis*, SPIE press Bellingham (2007).
13. B. Hermann, E. Fernández, A. Unterhuber, H. Sattmann, A. Fercher, W. Drexler, P. Prieto and P. Artal, "Adaptive-optics ultrahigh-resolution optical coherence tomography," *Optics letters* **29**(18), 2142-2144 (2004)
14. R. Zawadzki, S. Jones, S. Olivier, M. Zhao, B. Bower, J. Izatt, S. Choi, S. Laut and J. Werner, "Adaptive-optics optical coherence tomography for high-resolution and high-speed 3D retinal in vivo imaging," *Optics express* **13**(21), 8532-8546 (2005)
15. M. Rueckel, J. A. Mack-Bucher and W. Denk, "Adaptive wavefront correction in two-photon microscopy using coherence-gated wavefront sensing," *Proceedings of the National Academy of Sciences* **103**(46), 17137-17142 (2006)
16. K. Kurokawa, K. Sasaki, S. Makita, M. Yamanari, B. Cense and Y. Yasuno, "Simultaneous high-resolution retinal imaging and high-penetration choroidal imaging by one-micrometer adaptive optics optical coherence tomography," *Optics express* **18**(8), 8515-8527 (2010)

17. N. Iftimia, B. E. Bouma and G. J. Tearney, "Speckle reduction in optical coherence tomography by "path length encoded" angular compounding," *J Biomed Opt* **8**(2), 260-263 (2003)
18. M. Pircher, E. Go, R. Leitgeb, A. F. Fercher and C. K. Hitzenberger, "Speckle reduction in optical coherence tomography by frequency compounding," *J Biomed Opt* **8**(3), 565-569 (2003)
19. M. Szkulmowski, I. Gorczynska, D. Szlag, M. Sylwestrzak, A. Kowalczyk and M. Wojtkowski, "Efficient reduction of speckle noise in Optical Coherence Tomography," *Optics Express* **20**(2), 1337-1359 (2012)
20. A. E. Desjardins, B. J. Vakoc, W. Y. Oh, S. M. R. Motaghiannezam, G. J. Tearney and B. E. Bouma, "Angle-resolved Optical Coherence Tomography with sequential angular selectivity for speckle reduction," *Optics Express* **15**(10), 6200-6209 (2007)
21. A. P. Mosk, A. Lagendijk, G. Lerosey and M. Fink, "Controlling waves in space and time for imaging and focusing in complex media," *Nature photonics* **6**(5), 283-292 (2012)
22. H. Yu, J. Park, K. Lee, J. Yoon, K. Kim, S. Lee and Y. Park, "Recent advances in wavefront shaping techniques for biomedical applications," *Current Applied Physics* **15**(5), 632-641 (2015)
23. I. M. Vellekoop and A. Mosk, "Focusing coherent light through opaque strongly scattering media," *Optics Letters* **32**(16), 2309-2311 (2007)
24. R. Fiolka, K. Si and M. Cui, "Complex wavefront corrections for deep tissue focusing using low coherence backscattered light," *Optics Express* **20**(15), 16532-16543 (2012)
25. H. Yu, J. Jang, J. Lim, J.-H. Park, W. Jang, J.-Y. Kim and Y. Park, "Depth-enhanced 2-D optical coherence tomography using complex wavefront shaping," *Optics Express* **22**(7), 7514-7523 (2014)
26. J. Jang, J. Lim, H. Yu, H. Choi, J. Ha, J. H. Park, W. Y. Oh, W. Jang, S. Lee and Y. Park, "Complex wavefront shaping for optimal depth-selective focusing in optical coherence tomography," *Optics Express* **21**(3), 2890-2902 (2013)
27. S. Kang, S. Jeong, W. Choi, H. Ko, T. D. Yang, J. H. Joo, J.-S. Lee, Y.-S. Lim, Q.-H. Park and W. Choi, "Imaging deep within a scattering medium using collective accumulation of single-scattered waves," *Nature Photonics* **9**(4), 253-258 (2015)
28. J. Yoon, K. Lee, J. Park and Y. Park, "Measuring optical transmission matrices by wavefront shaping," *Optics Express* **23**(8), 10158-10167 (2015)
29. H. Yu, J.-H. Park and Y. Park, "Measuring large optical reflection matrices of turbid media," *Optics Communications* **352**(33-38 (2015)
30. H. Yu, T. R. Hillman, W. Choi, J. O. Lee, M. S. Feld, R. R. Dasari and Y. Park, "Measuring Large Optical Transmission Matrices of Disordered Media," *Phys Rev Lett* **111**(15), 153902 (2013)
31. S. A. Prahl, M. J. van Gemert and A. J. Welch, "Determining the optical properties of turbid mediaby using the adding–doubling method," *Applied optics* **32**(4), 559-568 (1993)
32. D. B. Conkey, A. M. Caravaca-Aguirre and R. Piestun, "High-speed scattering medium characterization with application to focusing light through turbid media," *Optics express* **20**(2), 1733-1740 (2012)
33. Y. Liu, P. X. Lai, C. Ma, X. Xu, A. A. Grabar and L. V. Wang, "Optical focusing deep inside dynamic scattering media with near-infrared time-reversed ultrasonically encoded (TRUE) light," *Nature Communications* **6**((2015)
34. M. Cui, E. J. McDowell and C. H. Yang, "An in vivo study of turbidity suppression by optical phase conjugation (TSOPC) on rabbit ear," *Optics Express* **18**(1), 25-30 (2010)
35. J.-H. Park, C. Park, H. Yu, Y.-H. Cho and Y. Park, "Dynamic active wave plate using random nanoparticles," *Opt. Exp.* **20**(15), 17010-17016 (2012)
36. J. H. Park, C. H. Park, H. Yu, Y. H. Cho and Y. K. Park, "Active spectral filtering through turbid media," *Opt. Lett.* **37**(15), 3261-3263 (2012)

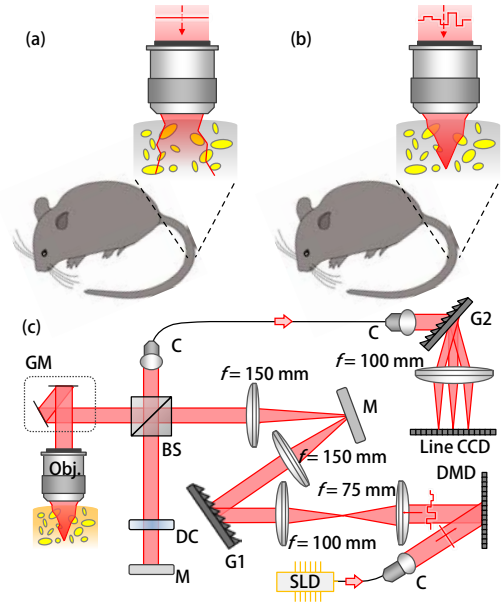

**Fig. 1** Experimental scheme for scattering-controlled *in vivo* tissue imaging. (a) Inhogeneous distributions of refractive indices in tissues cause multiple light scatterings, which prevent light delivery to a specific point. (b) By controlling an impinging wavefront, light can be constructively interfered with at the target position to allow depth-enhanced OCT imaging. (c) Schematic of the experimental setup. C, collimator; M: mirror, G1-2: gratings, BS: beam splitter, GM: galvanometric mirror, DC: dispersion compensator

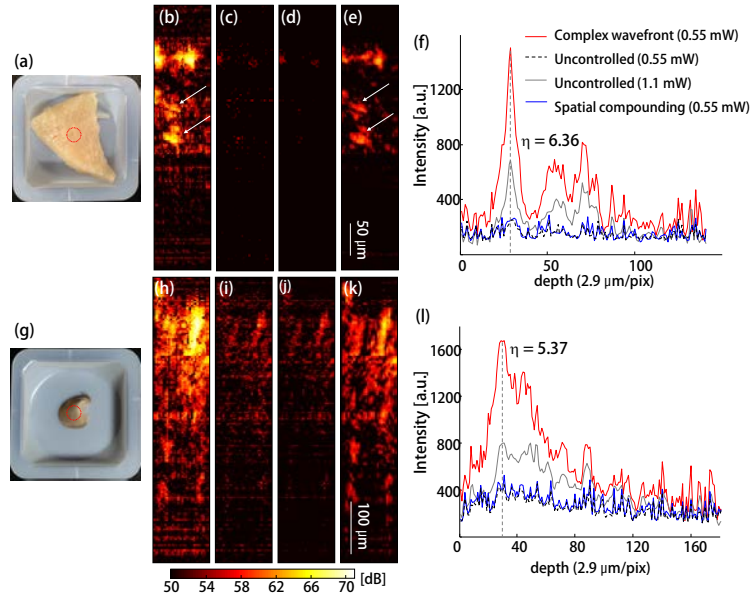

**Fig. 2** *In vitro* WS-OCT images of the chicken breast (a-f) and mouse ear tissue (g-l). Photographs of the chicken breast (a) and mouse ear tissue (g). The image areas are indicated as the red dashed circles. (b,h) Optimized image by wavefront control with input power of 0.55 mW. (c,i) Image acquired with an uncontrolled input beam of 0.55 mW. (d,j) Image obtained with the spatial compounding method. (e,k) Image acquired with an uncontrolled input beam of 1.1 mW. (f,l) The averaged A-scan profiles along 15 different A-scans in each case were plotted for the chicken breast and the mouse ear tissue, respectively.

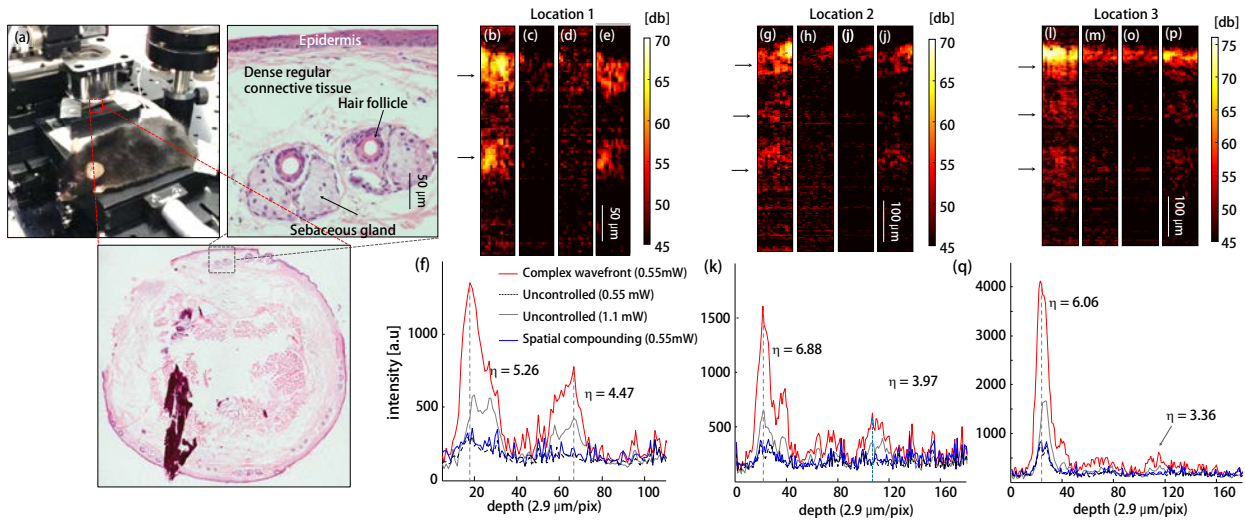

**Fig. 3** *In vivo* WS-OST images of the tissue in a tail of a live mouse. (a) A mouse under imaging and optical micrographs of a tail section slice of 30 μm, measured with a bright-field microscopy after H&E staining. (b-q) OCT images were acquired at three different locations. (b,g,l) Optimized image by wavefront control with an input power of 0.55 mW. Multilayered structrues are indicated as the gray arrows. The shapes and locations of these structures corresponds to epidermis, connective tissue textures, and sebaceous glands. (c,h,m) Image acquired with an uncontrolled input beam of 0.55 mW. (d,i,o) Image obtained with the spatial compounding method. (e,j,p) Image acquired with an uncontrolled input beam of 1.1 mW. (f,k,q) The averaged A-scan profiles along 15 different A-scans.

7